\documentclass{pnastwo}

\usepackage[dvips]{graphicx}

\usepackage{amssymb,amsfonts,amsmath}

\contributor{Submitted to Proceedings
of the National Academy of Sciences of the United States of America}
\url{www.pnas.org/cgi/doi/10.1073/pnas.0709640104}
\copyrightyear{2008}
\issuedate{Issue Date}
\volume{Volume}
\issuenumber{Issue Number}

\begin{document}



\title{A two-stage algorithm for extracting the multiscale backbone of complex weighted networks}





\author{Paul B. Slater\affil{1}{University of California, Santa Barbara}}

\contributor{Submitted to Proceedings of the National Academy of Sciences
of the United States of America}

\maketitle

\begin{article}




The central problem of concern to
Serrano, Bogu\~na and Vespignani \cite{SBV} can be 
effectively and elegantly addressed using a well-established two-stage algorithm 
that has been applied to internal
migration flows for numerous nations and several 
other forms of "transaction flow
data" \cite{japan,france,hubsclusters}. 

In the first stage, the $N$ row and $N$ column sums of the $N \times N$ matrix of weighted, directed network flows ($f_{ij}$) are alternately scaled to all equal 1, iteratively, until sufficient convergence to a {\it doubly}-stochastic table is attained. So doing serves as a control for multiscale effects.
(This procedure--convergent under broad conditions \cite{hartfiel}--leaves the cross-product ratios 
$\frac{f_{ij} f_{kl}}{f_{il} f_{kj}}$ [measures of association] invariant, and provides maximum-entropy estimates given the doubly-stochastic constraint. One may also consider for highly sparse networks, a preliminary smoothing of the 
matrix entries, or an adjustment of row and column sums to be proportional, not to 1, but to the number of non-zero entries in the row or column. Powering a doubly-stochastic matrix yields another [smoother] doubly-stochastic matrix.) 

In the second stage of the algorithm, 
the $N$ nodes (vertices) of the network are {\it hierarchically} clustered. One starts with an $N$-node graph with no links. Then, if the $ij$-entry 
of the doubly-standardized table is the greatest, a link is 
drawn from node $i$ to node $j$, and so on, using the second, third...
greatest links. Larger and larger strong components of the directed graph (digraph) emerge, until all the nodes are united in a single strong component (that is, there exists a path of directed links from any node to any other). This final digraph (with links labeled by the original flow values) is an obvious candidate for the "multiscale backbone" of the complex weighted network. (In our recent study of migration between the 3,107 U. S. 
counties, this backbone consisted of 25,329 links \cite{hubsclusters}. 
In the methodology of Serrano, Bogu\~na and Vespignani \cite{SBV}, some of the $N$ nodes may be omitted from the backbone, depending upon the significance level $\alpha$ employed, while the {\it single} strong component includes all $N$ nodes. By stopping the hierachical clustering before completion, however, in some suitable statistical manner, one might also isolate nodes.) An $O(M \log{N})$ algorithm, where $M$ is the number of edges of the network--developed by R. E. Tarjan--exists for the hierarchical clustering. If the digraph of the original network is not strongly connected, the sub-networks corresponding to its individual strong components can each be independently analyzed using the two-stage algorithm \cite{hartfiel}.

The two-stage procedure has proved effective--to use the internal migration context--in recognizing regional structures (such as the islands of 
Japan \cite{japan} and French Louisiana \cite{hubsclusters}) and in distinguishing
"cosmopolitan" geographic units (such as Paris \cite{france} and U. S. "Sunbelt" counties \cite{hubsclusters})--those with broad (hub-like 
or centralized) ties--from "provincial" units  \cite{japan,france,hubsclusters}. Graph-theoretic tests of an ordinal nature can be applied to assess the significance of the clusters found \cite[sec. IV.B]{hubsclusters}.








\end{article}



\end{document}